\documentclass[secnumarabic,twocolumn]{revtex4-1}

\usepackage{amsmath,amssymb,amsthm}

\usepackage{hyperref}
\usepackage{amsthm}
\usepackage{graphicx}
\usepackage{subfigure}
\usepackage{color}
\usepackage[normalem]{ulem}

\definecolor{violet}{rgb}{1.00,0.00,1.00}	% violet
\definecolor{orange}{rgb}{1.00,0.50,0.00}	% orange
\definecolor{turq}{rgb}{0.00,0.6,1.00}		% turquoise

\theoremstyle{plain}

\begin{document}

% \title{A low-level form of sensory awareness in asynchronous network states}

\title{Enhanced responsiveness in asynchronous irregular neuronal networks}

% \author{Yann Zerlaut$^{1}$,

\author{Zahara Girones$^{1}$ and Alain Destexhe$^{1,2,3}$\\
  {\small 1. Unit for Neurosciences, Information and Complexity
    (UNIC), CNRS, Gif sur Yvette, France, \\
    2. The European Institute for Theoretical Neuroscience (EITN), Paris \\
    3. Correspondence to destexhe@unic.cnrs-gif.fr\\
  }}

\date{\today}

\begin{abstract}

  Networks of excitatory and inhibitory neurons display asynchronous
  irregular (AI) states, where the activities of the two populations
  are balanced.  At the single cell level, it was shown that neurons
  subject to balanced and noisy synaptic inputs can display enhanced
  responsiveness.  We show here that this enhanced responsiveness is
  also present at the network level, but only when single neurons are
  in a conductance state and fluctuation regime consistent with
  experimental measurements.  In such states, the entire population of
  neurons is globally influenced by the external input.  We suggest
  that this network-level enhanced responsiveness constitute a
  low-level form of sensory awareness.

\end{abstract}

\pacs{87.10.+e, 05.40.+j, 02.50.Fz}

\maketitle

Networks of neurons are capable of displaying asynchronous and
irregular (AI) states of activity, where neurons fire in an apparent
stochastic fashion, and with very low levels of synchrony.  This
property was found in networks of excitatory and inhibitory neurons,
in a balanced state, first by binary neurons~\cite{VV96,VV98}, and
subsequently by more complex models such as spiking
neurons~\cite{Amit97,Brunel2000,Vogels2005,Destexhe2009,Yger2011}.

In parallel, a characterization of the noisy background activity seen
in cortical neurons {\it in vivo}~\cite{Pare98} showed that these
neurons are in a high-conductance (HC) state~\cite{DP99,Destexhe2003}.
HC states were first shown {\it in vivo} under anesthesia, then
confirmed in the awake brain~\cite{Steriade2001,Rudolph2007,Petersen}.
HC states were also largely investigated {\it in vitro} by a number of
studies using the dynamic-clamp technique.  These studies have
collectively shown that the HC state confers to neurons enhanced
responsiveness, which is due to a tight interplay between conductances
and $V_m$
fluctuations~\cite{Destexhe2001,Chance2002,Rauch,Wolfart2005,Prescott,Zerlaut2016}.

In the present paper, we attempt to unify these two points of view by
showing that AI states in networks can display enhanced responsiveness
properties, but only if neurons display the correct conductance state
and fluctuation regime.

\begin{figure}[htbp]
       \centering \includegraphics[width=0.9\columnwidth]{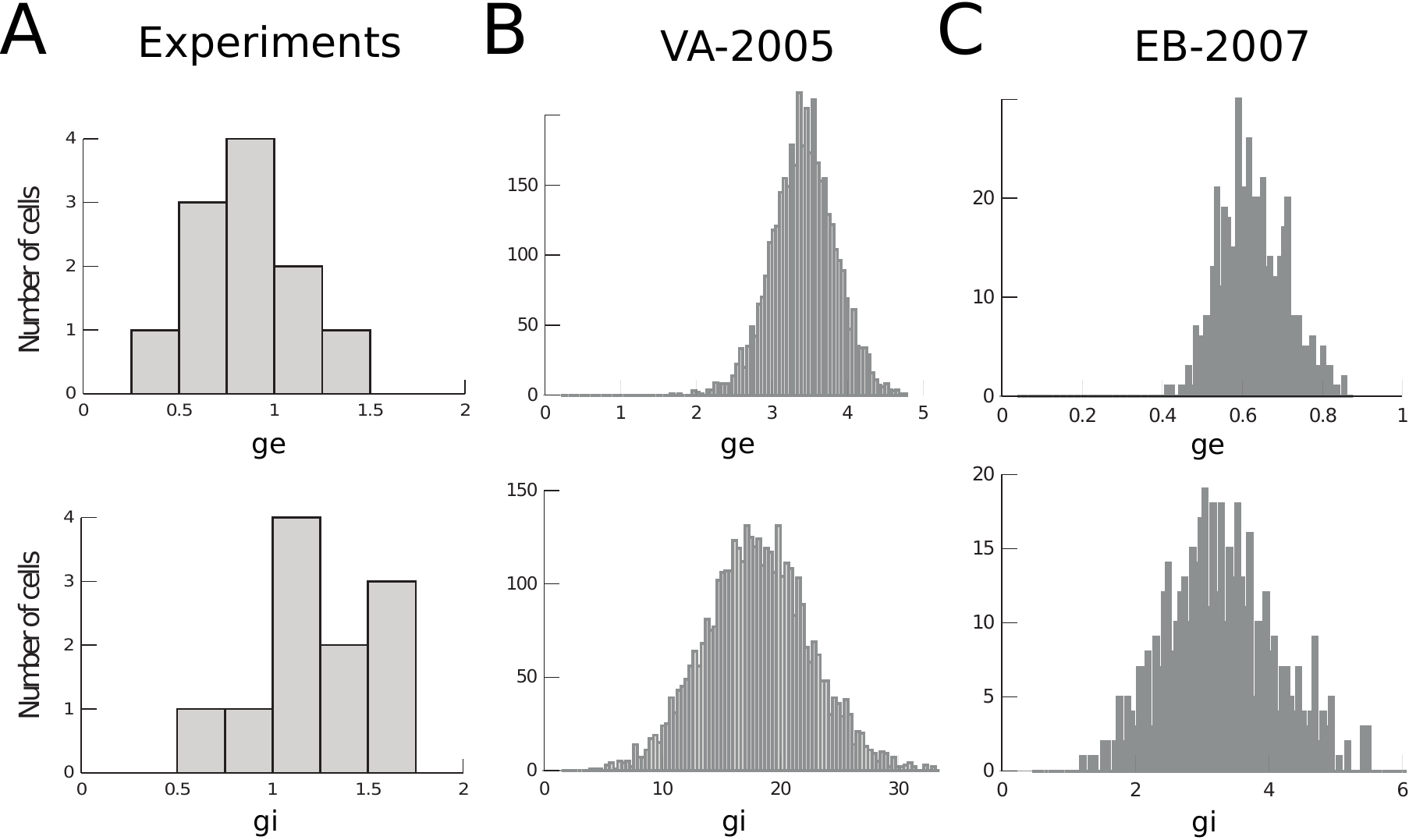}

       \caption{ {\bf Conductance state of different networks}.  A.
         Conductance measurements in cat cortical neurons during
         wakefulness.  The distributions represent the total
         excitatory and inhibitory conductances, divided by the leak
         conductance of the cell ($g_e$ and $g_i$, respectively; data
         from Rudolph et al.\ (2007).  B.  Conductance distribution in
         a network of leaky integrate and fire neurons displaying AI
         states.  The model (VA-2005) is taken from Vogels and Abbott
         (2005).  C.  Conductance distributions for another network
         model of AI states (EB-2007), taken from El Boustani et al.\
         (2007).}

\label{VA}
\end{figure}

Figure~\ref{VA} compares different network models displaying AI
states, with experimental measurements.  The measurements of
excitatory and inhibitory conductances in neurons is of primary
importance, as we will show here.  Conductance measurements in
cortical neurons in awake cats are depicted in Fig.~\ref{VA}A.  There
was a very wide range of conductance values measured from cell to
cell, but when normalized to the estimated resting conductance,
approximately symmetric distributions were
obtained~\cite{Rudolph2007,NeuroComp2007}.  The total excitatory
conductance is slightly lower than the resting conductance, while
inhibitory conductances are in general about 1.5 times larger than the
leak (Fig.~\ref{VA}A).

% wake analysis
% VA raster  (color = Raster2)
% conductance analysis  (Abbott\_analysis)
% new raster
% new conductance analysis  (LCstate)

These measured conductances were compared to a well-known model of AI
state consisting of 5000 randomly-connected leaky integrate and fire
neurons (80\% excitatory, 20\% inhibitory)~\cite{Vogels2005}.  We
computed the total excitatory and inhibitory conductances seen in
different neurons of this model.  It appeared that, in this
5000-neuron network, when normalized to the resting conductance, the
relative conductances in this model were considerably larger than
physiological measurements, up to about 20 times larger
(Fig.~\ref{VA}B).  Another model of AI state~\cite{NeuroComp2007},
consisting of 16,000 randomly-conected neurons, but with smaller
synaptic weights, did not display such an aberrant conductance state,
and could generate a conductance state much closer to physiological
measurements~\cite{NeuroComp2007} (Fig.~\ref{VA}C).

\begin{figure}[htbp]
       \centering \includegraphics[width=0.8\columnwidth]{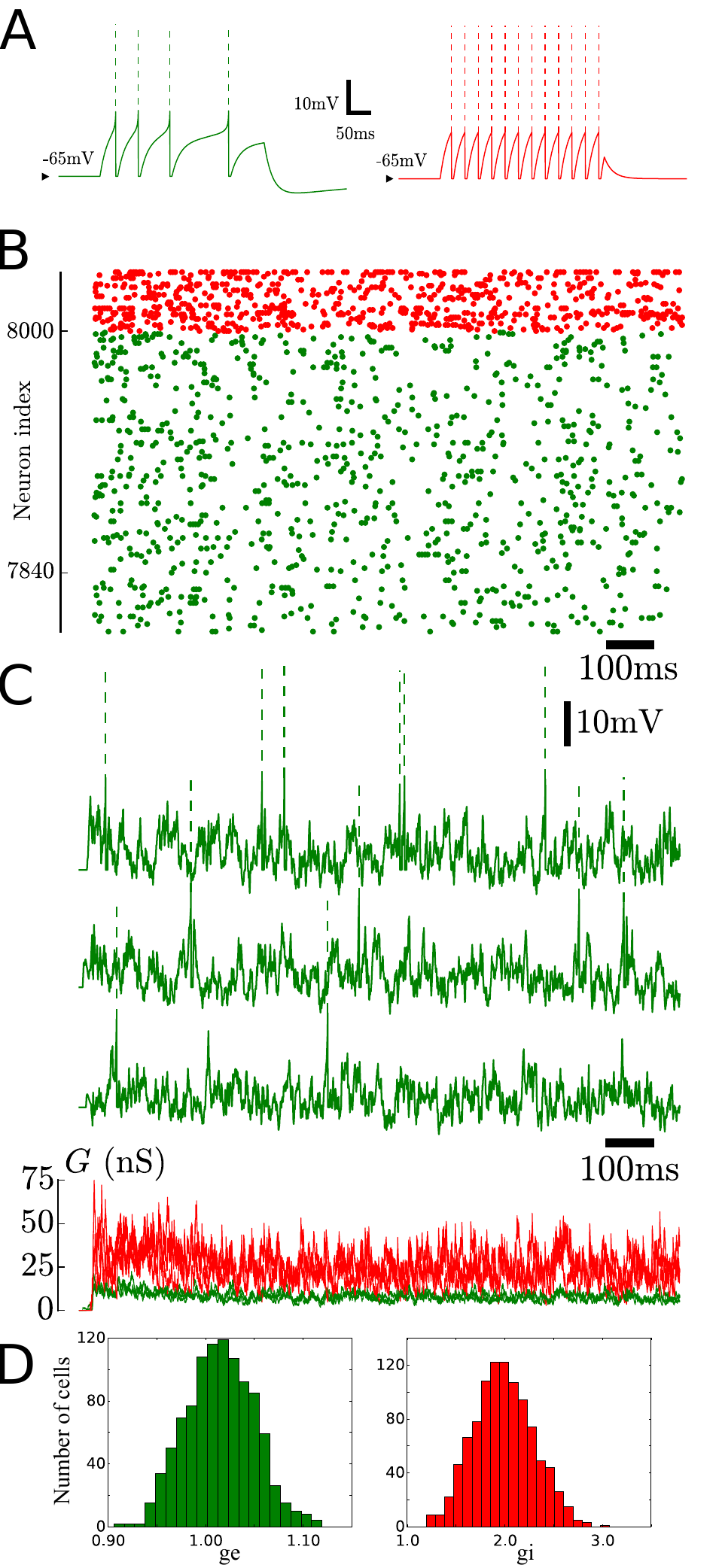}

       \caption{ {\bf Biophysically realistic network model of AI
           states}.  (color online) A. Two cell types used in the
         model, respectively for excitatory (RS cell, green) and
         inhibitory (FS cell, red).  The traces show typical responses
         to depolarizing current pulses.  B.  Raster of activity
         during an AI state generated by this network.  C. Examples of
         excitatory cells (top, green traces) and their total synaptic
         conductance (red inhibitory, green excitatory). D.
         Conductance distribution in this network, normalized to the
         leak conductance.  Parameters are given in Appendix.}

\label{rsfs}
\end{figure}

Such a physiologically plausible conductance state was also obtained
in a more realistic network of excitatory and inhibitory cells, where
the adaptation of excitatory cells was taken into account
(Fig.~\ref{rsfs}).  This model used two types of cells, the ``regular
spiking'' (RS) and ``fast spiking'' (FS) neurons, which correspons to
the typical firing patterns seen {\it in vitro}~\cite{McCormick85}.
These cell types were modeled using the Adaptive Exponential integrate
and fire model~\cite{Brette2005} (Fig.~\ref{rsfs}A; see details in
Appendix).  A network of 8000 RS and 2000 FS cells generated AI states
with the typical higher frequency firing of inhibitory neurons
(Fig.~\ref{rsfs}B).  In this network, the conductance state of
individual cells was consistent with experimental measurements in
awake animals (Fig.~\ref{rsfs}C). The conductance distributions
obtained in this network model (Fig.~\ref{rsfs}D) are close to
experimental values (compare with (Fig.~\ref{VA}A).

\begin{figure}[htbp]
       \centering \includegraphics[width=\columnwidth]{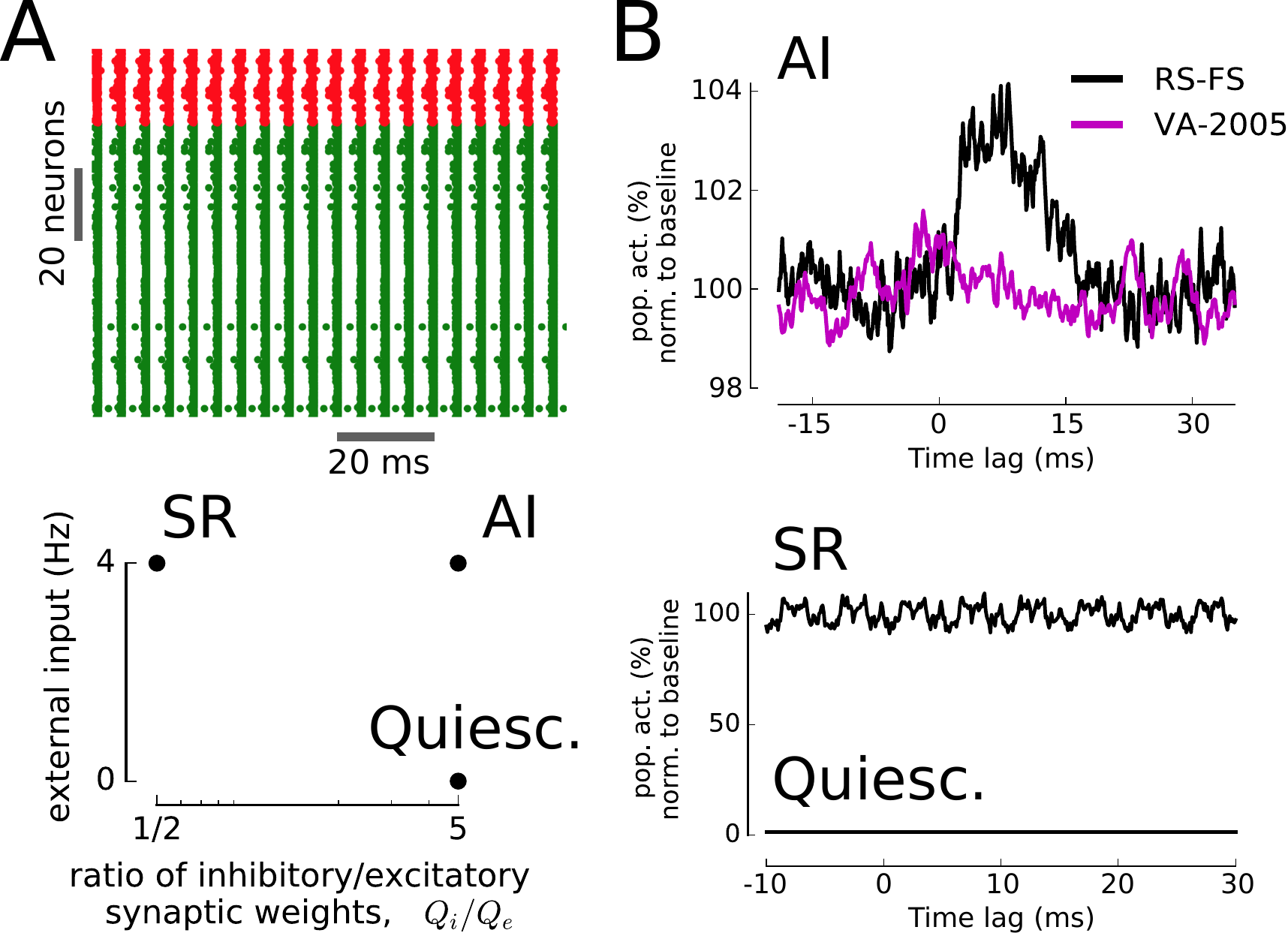}

       \caption{ {\bf Enhanced responsiveness in AI states at the
           network level}. (color online) A. Three different states in
         the parameter space of the RS-FS cells network, synchronous
         regular (SR), asynchronous irregular (AI) and quiescent
         (Quiesc).  B.  Responsiveness of the network according to its
         activity state.  The average population response is shown for
         an excitatory input randomly distributed on a subset of cells
         (1 single EPSP of 1nS distributed in 40 cells; average of
         4000 trials).  The response in another
         network~\cite{Vogels2005} (VA-2005), with aberrant
         conductance state, is shown for comparison (magenta).}

\label{enhAI}
\end{figure}

% yann raster    (from thesis)
% conductance analysis
% show responsiveness   (edit figure)

Interestingly, comparing these networks in term of their
responsiveness, revealed that the conductance state of the network is
very important.  When an external input was given to a randomly chosen
sample of cells, the network in a physiologically plausible AI state
could display a remarkable sensitivity to this external input
(Fig.~\ref{enhAI}).  However, in a network displaying an aberrant
conductance state, this response was not present.  The response was
also absent in a quiescent state, or in an oscillatory state in the
same network (Fig.~\ref{enhAI}B).  This shows that the AI state, if
with correct conductance and $V_m$ fluctuations, displays an enhanced
responsiveness to external inputs, and is able to collectively detect
inputs of very small amplitude, which normally would have been
subthreshold.

\begin{figure}[htbp]
       \centering \includegraphics[width=0.8\columnwidth]{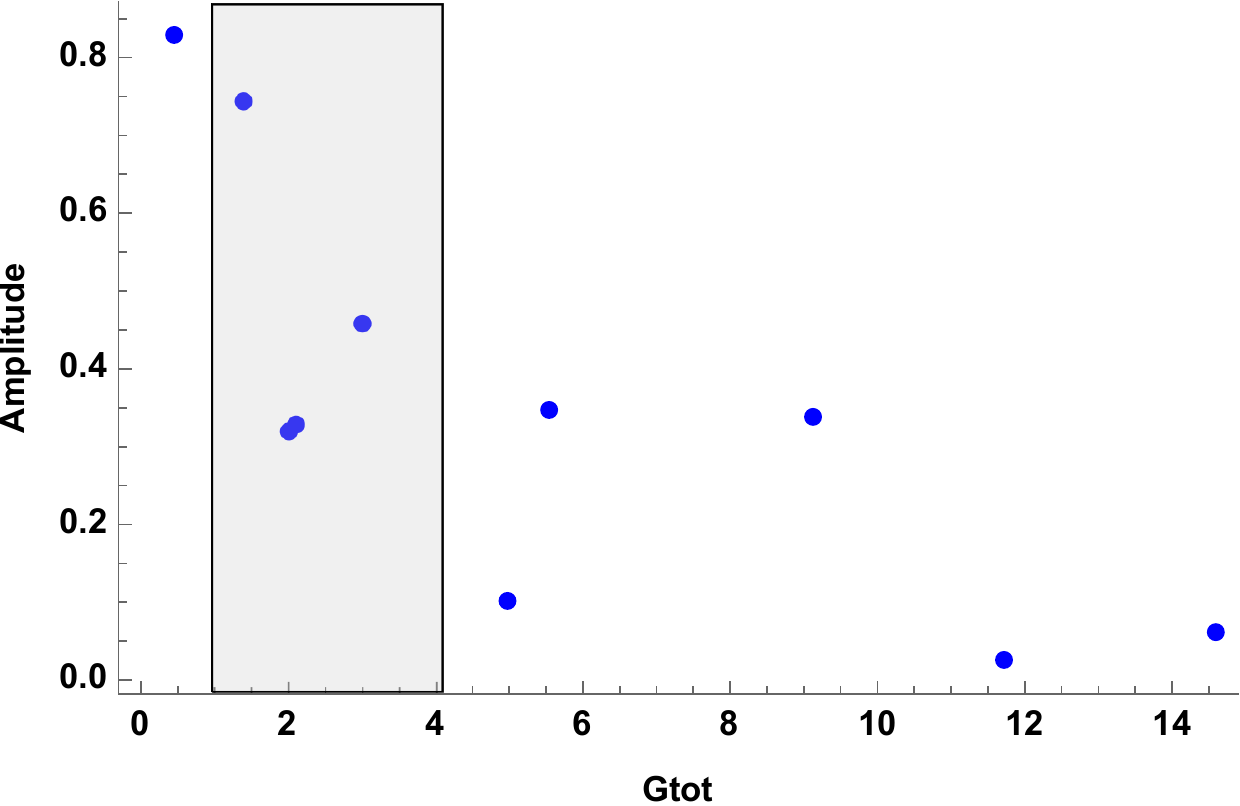}

       \caption{ {\bf Ehanced responsiveness depends on conductance state.}
         (color online) The average response is represented against the
         total conductance, for randomly selected parameter sets displaying
         AI states.  The response was computed as the integral of the
         population response (as in Fig.~\ref{enhAI}B), above the baseline.
         The total conductance was calculated as the sum of the average
         conductance distributions (as in Fig.~\ref{rsfs}D), and thus was
         normalized to the leak conductance.  The gray area indicates the
         conductance range corresponding to physiological measurements in
         awake animals.}

\label{rob}
\end{figure}

We have done additional simulations to explore how responsiveness
depends on conductance state, by considering random sets of parameters
around the AI state in Fig.~\ref{enhAI}A.  This exploration showed
that the responsiveness highly depends on the conductance state of the
network (Fig.~\ref{rob}).  The response was quantified as the area of
the evoked population response (as in Fig.~\ref{enhAI}B), represented
against the total conductance, as calculated from distributions (as in
Fig.~\ref{rsfs}D).  One can see that the response is indeed high for
physiological conductance states (gray area in Fig.~\ref{rob}) and
approaches to zero for aberrant conductance states.  This result
explains the difference of responsiveness between the two networks
shown in Fig.~\ref{enhAI}B (top).  However, it is important to keep in
mind that the response also depends on the level of fluctuations and
the average V$_m$ level, which can be quantified by considering
distributions of the standard deviation of the membrane potential
($\sigma_V$) and of the mean V$_m$ of the cells (not shown).  The
responsiveness can also be understood using a phenomenological model,
as shown recently~\cite{Reig2015}.

% cite here the perspective paper ...

This phenomenon of network-level responsiveness is very similar to the
enhanced responsiveness that was found earlier at the single neuron
level~\cite{Ho2000}.  In single-cell studies, the presence of synaptic
``noise'' conferred an anhanced sensitivity to the neuron.  It was
also shown that having the correct conductance state and fluctuation
regime of the cell is important, it sets the response to a realistic
level, and enhanced responsiveness was present using the conductances
measured experimentally, as well as their level of fluctuations.
Enhanced responsiveness was also found in noisy networks, either in
feedforward networks where all cells display a high-conductance
state~\cite{Ho2000}, or in chaotic recurrent networks where the
enhanced responsiveness was quantified in terms of information
transport~\cite{Destexhe94,Destexhe-Contreras2006}.  One of these
studies was the first to demonstrate that networks in a chaotic state
display enhanced responsiveness~\cite{Destexhe94}.  We complement here
these previous studies by showing that enhanced responsiveness is
present in recurrent networks displayin AI states, which are
presumably chaotic~\cite{VV98,VV98}.  In addition, we show that this
property highly depends on conductance state, and that physiologically
plausible conductance state are particularly responsive.  This
underlies the importance to work with conductance-based models, as
current-based models are unlikely to display the correct
responsiveness, and cannot be checked against conductance
measurements, so are also unconstrained.

Thus, the present results suggest that the conductance state of a
network is a fundamental property to understand its responsiveness,
which emphasize the importance of conductance measurements {\it in
  vivo}.  We suggest that HC states have a universal aspect, in the
sense that the conductance and fluctuation level measured {\it in
  vivo} are the fundamental parameters that neural networks should
reproduce, to yield responsiveness properties relevant to neural
function.

Finally, let us emphasize that in the AI state, the network responds
instantaneously to an input, and this input can occur at any time,
which presents evident useful computational properties.  A
particularly interesting property is that, when submitted to
successive presentations of the same stimulus, different neurons
respond on each trial, showing that it is the whole network that is
globally ``aware'' of this stimulus.  We therefore propose that such
network-level responsiveness during AI states implements a low-level
form of sensory awareness in networks of neurons.  The fact that this
occurs in AI states is consistent with the observation that cerebral
cortex is in a ``desynchronized'' state in awake and attentive states
(reviewed in~\cite{Niedermeyer,Steriade-book}).  A recent review
concluded that desynchonized cortical activity is so far the best
correlate of conscious states~\cite{Koch-Nat-Rev}, but no mechanism
was given.  We propose here such a possible mechanism to link these
high-level aspects to elementary biophysical properties of neurons.

\medskip 

\noindent {\bf Acknowledgments: } Research funded by the CNRS and
the European Community ({\it Human Brain Project} H2020-720270). 
We thank Yann Zerlaut for many discussions and help with the
simulations.

% Y.Z. was supported by fellowships from the {\it Initiative
% d'Excellence Paris-Saclay} and the {\it Fondation pour la
% Recherche M\'edicale} (FDT 20150532751).

%\clearpage

\section*{Appendix}

\subsection*{Details of the network model}

The network model used here consisted of excitatory (RS) and
inhibitory (FS) neurons described by the Adaptive Exponential
integrate and fire model \cite{Brette2005}, for which the equations
for the membrane potential and the adaptation current read:
\begin{equation}
\label{eq:iAdExp}
  \left\{
  \begin{split}
  & C_m\,\frac{dV}{dt} = g_{L} \,(E_{L}-V) + I_{syn}(V,t) + k_a e^{\frac{V - V_{thre} }{k_a}}- I_w \\
  & \tau_w \frac{d I_w}{dt} = - I_w + \sum_{t_s \in \{t_{spike}\}} b \, \, \delta (t-t_s)
  \end{split}
\right. ~ ,
\nonumber
\end{equation}
where $V$ is the membrane potential, $C_m$ is the membrane
capacitance, $g_{L}$ is the resting conductance and $E_{L}$ its
reversal potential, $I_{syn}$ is the synaptic current, $k_a$ and
$V_{thre}$ are threshold parameters.  $I_w$ is the adaptation current,
which evolved according to a time constant $\tau_w$ and increases by
a value $b$ at each spike $t_s$.  FS cells correspond to the same model
with $I_w$=0.

The cellular parameters were, for RS cells: $g_L$=10nS, $C_m$=200pF,
$T_{ref}$=5ms, $E_L$=-70mV, $V_{thre}$=-50mV, $V_{reset}$=-70mV,
$k_a$=2mV, $a$=4nS, $b$=20pA, $\tau_w$=500ms.  FS cells were descrivbed
by a leaky integrate and fire model with $g_L$=10nS, $C_m$=200pF,
$T_{ref}$=5ms, $E_L$=-70mV, $V_{thre}$=-50mV, $V_{reset}$=-70mV.
Network parameters: 4000 RS and 1000 FS cells, connectd randomly with
probability of 5\%, and with synaptic weights of $Q_e$=1nS and
$Q_i$=5nS (with respective reversal potentials of $E_e$=0mV,
$E_i$=-80mV). The synaptic conductances were decaying exponentials
with a time constant of 5~ms.

An external drive was present in all neurons and consisted of 4000
independent Poisson processes (rate of 4~Hz), projected over the 5000
neurons with 5\% connection probability (weight of 1nS).

The external stimulus of Fig~\ref{enhAI} consisted of 40 synchronous
EPSPs of 1nS, spread over 40 randomly-choosen excitatory neurons
within the population.

\end{document}